\documentclass[10pt]{iopart}
\pdfoutput=1
\usepackage{amssymb}
\usepackage{breqn}
\usepackage{multirow}
\usepackage[pdftex]{graphicx}
\usepackage[table]{xcolor}
\usepackage[flushleft]{threeparttable}
\DeclareGraphicsExtensions{.pdf,.jpeg,.png}
\begin{document}

\title{Modelling the Nonlinear Response of Fibre-reinforced Bending Fluidic Actuators}
\author{Vito Cacucciolo$^1$, Federico Renda$^2$, Ernesto Poccia$^1$, Cecilia Laschi$^1$, Matteo Cianchetti$^1$}

\address{$^1$ The BioRobotics Institute, Scuola Superiore Sant’Anna, Pisa, Italy}

\address{$^2$ Khalifa University Robotics Inbstitute, Khalifa University, Abu Dhabi, UAE}

\ead{v.cacucciolo@sssup.it}
\vspace{10pt}
%\begin{indented}
%\item[]February 2014
%\end{indented}
%

\begin{abstract}
Soft actuators are receiving increasing attention from the engineering community, not only in research but even for industrial applications. Among soft actuators, fibre-reinforced Bending Fluidic Actuators (BFAs) became very popular thanks to features such as robustness and easy design and fabrication. However, an accurate modelling of these smart structures, taking into account all the nonlinearities involved, is a challenging task. 
% Some FEM software can be used for simulation, but are often computationally inefficient and miss to capture the phenomena.
In this effort, we propose an analytical mechanical model to capture the quasi-static response of fibre-reinforced BFAs. The model is fully 3D and for the first time includes the effect of the pressure on the lateral surface of the chamber as well as the non-constant torque produced by the pressure at the tip. The presented model can be used for design and control, while providing information about the mechanics of these complex actuators.
\end{abstract}

\vspace{2pc}

% \submitto{\SMS}
%\maketitle
\ioptwocol
%

\section{Introduction}
\label{sec:intro}
Actuators made by compliant materials are attracting high attention in the recent years \cite{maeda2015rapid, shintake2016versatile, Bufalo2008, Carpi2013}. On one hand, soft robotics is proposing a novel approach to build service robots, meaning robots aimed at interacting with humans in their environments \cite{kim2013soft, laschi2014soft, majidi2014soft}. On the other, soft actuators and sensors find application also in wearable devices thanks to their intrinsic compliance, which makes them interact safely with the human body \cite{Carpi2013, Polygerinos2014, ranzani2015bioinspired}.
Among all the classes of soft actuators, flexible fluidic actuators (FFA) are one of the most widespread. FFAs are usually composed by a hollow structure made in hyper-elastic materials (e.g., silicone rubber) which inflates when compressed fluid is forced into the chamber.
We can distinguish two different categories of FFAs based on their actuation modes, the ones that deform axially and the ones that bend.

%McKibben models: (kinematics) \cite{Doumit2009}, \cite{bishop-moser2015design} (dynamic) \cite{Sorge2015}.

The most common example in the first category are McKibben actuators \cite{Doumit2009}. They are composed by a cylindrical ladder, where the compressed fluid (usually air) is inserted, and an external braid, composed by fibers arranged as spirals around the external surface of the cylinder. The consequence of inflation in McKibben actuators is both an axial displacement and a variation of the section. By designing the angle of the fibres it is possible to switch from axial elongation to contraction.

Bending fluidic actuators (BFAs), on the other hand, are slender structures that bend as a consequence of increase in internal pressure. Recently, they started to become widely used, finding applications in legged robots \cite{tolley2014resilient}, \cite{cacucciolo2015adaptive}, artificial hands \cite{deimel2014novel} and wearable rehabilitation devices \cite{Polygerinos2014}. They feature high robustness and easiness in fabrication. An additional advantage is that they fuse structure and actuator in one single element with infinite degrees of freedom but only one degree of actuation (the internal pressure).
BFAs can be split again into two groups on the base of the mechanism that transforms the expansion of the chamber into a bending of the whole structure. In the first group the geometry of the chamber is the main element driving the deformation \cite{Martinez2013}, \cite{Mosadegh2014}, while in the second we have fibre-reinforced bending actuators.

\begin{figure}[h]
\centering
\includegraphics[scale=1.0]{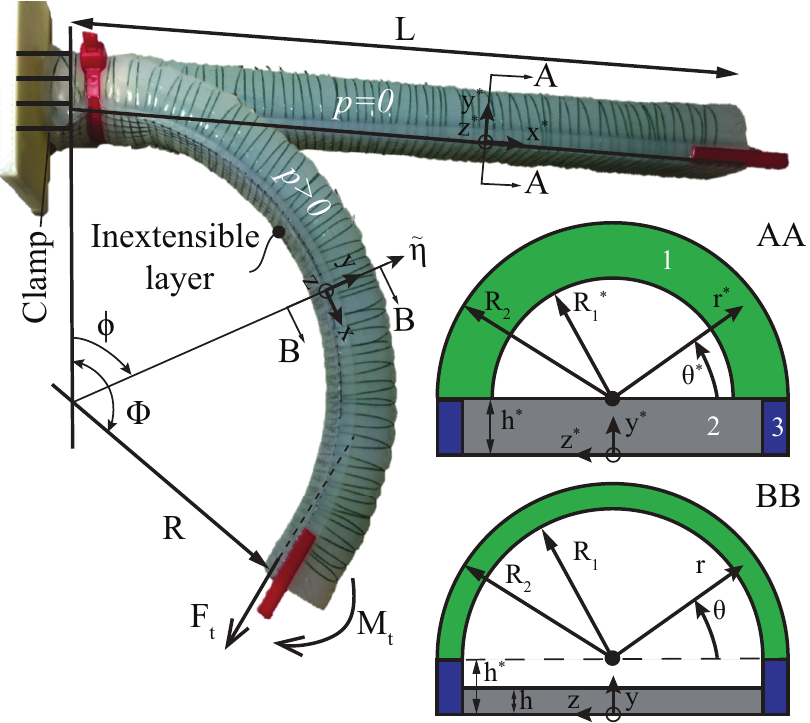}
\caption{Description of the fibre-reinforced Bending Fluidic Actuator under study with the schematics of the cross section. When pressure is applied $p>0$, the structure bends due to the elongation of the upper portion, while the cross section of the internal chamber expands.}
\label{fig:seg1}
\end{figure}

Fibre-reinforced BFAs can be though as McKibbens with the angle of the fibres $\gamma$ close to $\pi/2$, such that the inflation will results in almost pure elongation and no variation in the external perimeter of the cross-section, and with an asymmetric axial constraint (figure \ref{fig:seg1} and \ref{fig:exp}). This constraint, usually a fabric layer glued on one side of the structure, transforms the axial elongation in pure bending with the neutral surface coincident with the fabric layer.

Following the wide adoption of fibre-reinforced BFAs, an accurate mechanical model able to capture their nonlinear response can be very useful to push forward on the adoption of this technology. 
Modelling soft structures \cite{renda2014dynamic} and soft actuators \cite{bishop-moser2015design}, \cite{Sorge2015} is a challenging research topic due to its multi-disciplinary nature (e.g., fluid-structure interaction) and to the numerous nonlinearities involved (e.g., hyper-elastic materials, large deformations, non-homogeneous properties).
Up to know, few models have been proposed in the state of the art. Polygerinos et al. \cite{polygerinos2015modeling} realized a semi-analytical model using a hyper-elastic material model (Neo-hookean). Analytical models have some advantages over FEM models, such as providing insights about the physics of the phenomenon and fast computation. However, the simplifications applied in \cite{polygerinos2015modeling} on the kinematics of the structure make the model practically 1D. The authors did not take into account the effect of the pressure on the lateral surface of the inner chamber, which in facts contributes in the bending due to the incompressibility of the silicone rubber. Finally, they reduced all the effects of the pressure on the tip to a lumped torque constant with the bending angle. The result of all of these simplifications is that their models misses to capture the rich nonlinear behaviour of these actuators.

% that their model produces an almost linear response in terms of bending angle versus pressure, missing to capture the rich nonlinear behaviour of these actuators.

In this work, we propose an analytical 3D model to reproduce the response of the actuator as a consequence of the increase in internal pressure, when bending in free state. The model is quasi-static but can allows to be expanded to dynamic in further studies.
In this model, we take into account the effect of the transverse pressure as well as the variable torque created by the pressure at the tip. Additionally, we analyse all the three principal strains involved in the deformation of the structure and their interactions.

We analysed a slender structure with a semi-circular hollow cross-section, as shown in figure \ref{fig:seg1}.
The choice of this cross-section is arbitrary (the proposed approach can be applied to different shapes as well) and follows the finding in \cite{polygerinos2015modeling} that this geometry presents a relatively low bending resistance.
In order to describe the complex nonlinearities presented by this problem, we decided to use the Eulerian description and apply all the equilibria in the deformed configuration \cite{ogden1997nonlinear}. This approach allowed us to assume a-priori the modes of deformation of the structure and as a consequence apply simplifications that kept the model at an acceptable level of complexity, which enabled the analytical solution of the differential equations involved. The use of numerical computation became necessary only to evaluate algebraic integrals and high degree polynomial equations.
As for the description of the kinematics, the use of a toroidal coordinate system allowed us to keep the state of stress principal.

We validated the model on experiments conducted on actuators with three different geometries. We could capture the nonlinear response of the structure and its variation with the design parameter. This result is extremely important because enables the design of the response curve, a clue feature for coupling the actuators with other systems (e.g., the human body) or for exploiting snap-through instabilities \cite{Overvelde2015} or self-stabilized dynamical cycles \cite{cacucciolo2015adaptive}, \cite{sprowitz2013towards}.
Additionally, we analysed the state of stress and strain in the cross section of the structure, which provides important information for design choices about the materials or the dimensions.

Two key features of our model are computational lightness and readability. All the kinematics can be computed once for all and stored in a database, while each step of the statics requires only few numerical integrations, which are computationally efficient. On a standard laptop and using MATLAB for the numerical computations, each step of the statics requires few hundreds of seconds.
As for the readability, all of our assumptions are physically motivated and verified a-posteriori. Their validity gives insights into the complex mechanics of these structures and can contribute to the design of novel kind of actuators.

The paper is organized as follows. We first present the design and fabrication of the actuators in section \ref{sec:design}. Then section \ref{sec:model} contains the derivation of the mathematical model with all the assumptions, which are verified in the validation in section \ref{sec:results}. In the same section we present the results, both in terms of characterization of the response and structural mechanics, and we discuss about their importance.
Finally section \ref{sec:conclusions} contains the conclusions.

\section{Design and Fabrication}
\label{sec:design}
%%
%\begin{table*}
%\centering
%\def\arraystretch{1.2}%  1 is the default, change whatever you need
%\caption{Parameters of the three mock-ups tested in the experiments}
%\label{tab:mockup} 
%\begin{tabular}{|c|c|c|c|c|c|c|}
%\hline
%Mock-up \cellcolor[gray]{0.9}	& 
%$\mu$ [kPa] \cellcolor[gray]{0.9}	&
% $h^*$ [mm]  \cellcolor[gray]{0.9} &
%  $R_2$ [mm] \cellcolor[gray]{0.9} &
%   $R_1^*$ [mm] \cellcolor[gray]{0.9} &
%    $k^*$ \cellcolor[gray]{0.9} &
%    $ L$ [mm] \cellcolor[gray]{0.9}
%\\
%\hline
%1 \cellcolor[gray]{0.9} & 92 & 7.0 & 8.0 & 4.0 & 0.50 & 118
%\\
%\hline
%2 \cellcolor[gray]{0.9} & 92 & 7.7 & 8.0 & 2.7 & 0.33 & 118
%\\
%\hline
%3 \cellcolor[gray]{0.9} & 92 & 9.1 & 8.0 & 1.6 & 0.20 & 118
%\\
%\hline
%\end{tabular}
%\end{table*}
%%
%
\begin{table*}
\begin{center}
\caption{\label{tab:mockup}Parameters of the three mock-ups tested in the experiments.}
\begin{tabular}{@{}|c|c|c|c|c|c|c|}
\hline
Mock-up \cellcolor[gray]{0.9}	& 
$\mu$ [kPa] \cellcolor[gray]{0.9}	&
 $h^*$ [mm]  \cellcolor[gray]{0.9} &
  $R_2$ [mm] \cellcolor[gray]{0.9} &
   $R_1^*$ [mm] \cellcolor[gray]{0.9} &
    $k^*$ \cellcolor[gray]{0.9} &
    $ L$ [mm] \cellcolor[gray]{0.9}
\\
\hline
1 \cellcolor[gray]{0.9} & 92 & 7.0 & 8.0 & 4.0 & 0.50 & 118
\\
\hline
2 \cellcolor[gray]{0.9} & 92 & 7.7 & 8.0 & 2.7 & 0.33 & 118
\\
\hline
3 \cellcolor[gray]{0.9} & 92 & 9.1 & 8.0 & 1.6 & 0.20 & 118
\\
\hline
\end{tabular}
\end{center}
\end{table*}

We produced three different mock-ups of the actuator in order to run the validation tests, whose parameters are contained in table \ref{tab:mockup}.
We chose the geometrical parameters to be compatible with the use of the actuators as legs of the soft robot FASTT \cite{cacucciolo2015adaptive} or as fingers of a glove with rehabilitation purposes \cite{Polygerinos2014}.
As for the material property $\mu$, we obtained it from a single value declared by the producer of the silicone, rather than by fitting the model over the experimental data.

\begin{figure}[]
\centering
\includegraphics[width=\columnwidth]{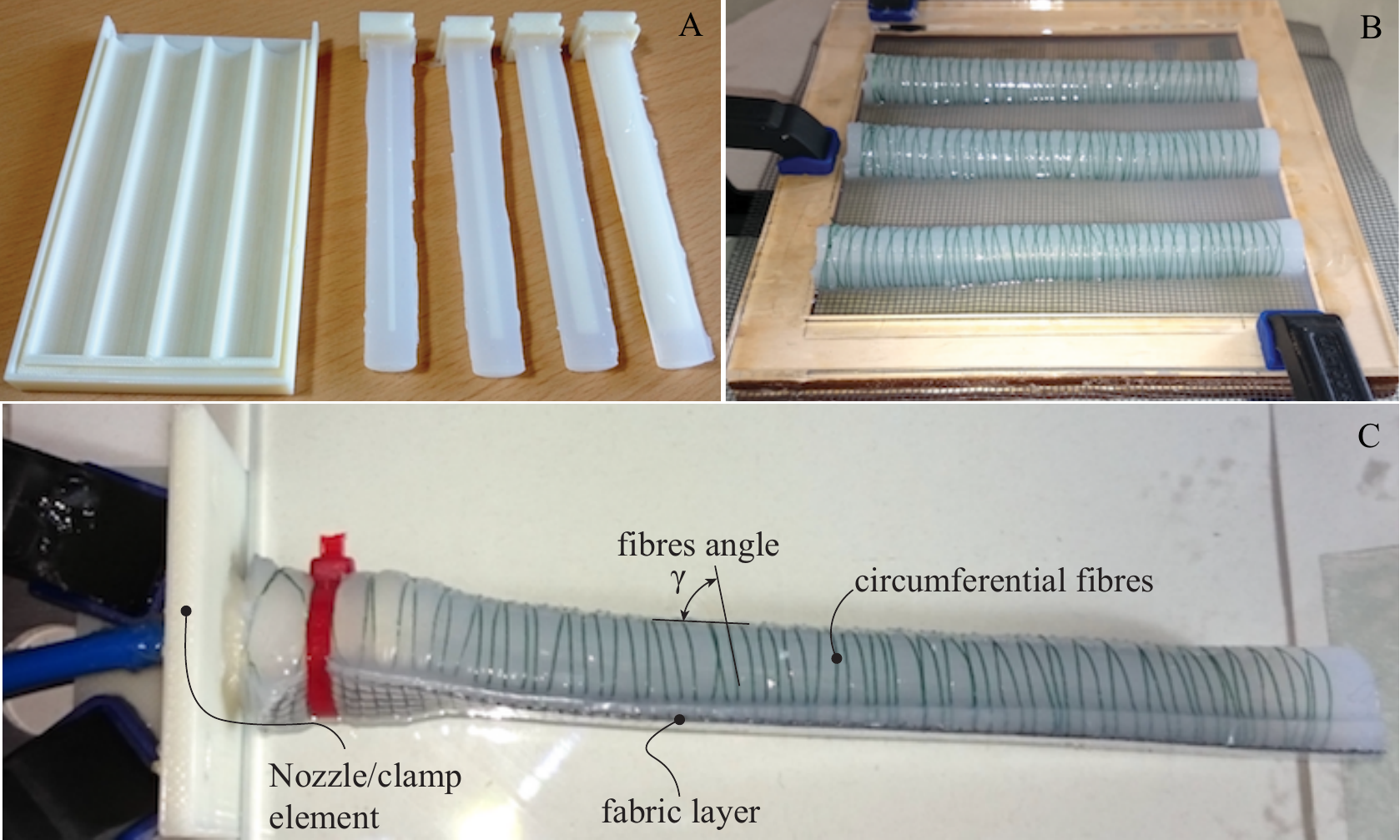}
\caption{Fabrication of the mock-ups. (A) Mold casting of the silicone bodies with inserts for the chambers; the obtained chambers are opened on one side only. (B) After the circumferential wrapping of the fibres, the fabric layer was incorporated by placing the mock-ups over a thin layer of un-cured silicone. (C) Each obtained structure was inserted on a custom 3D printed nozzle that acts as air inlets as well as mechanical clamp. The external diameter of the nozzle is bigger than the internal diameter of the chamber, in order to seal the chamber using the deformation of the silicone itself. With the aim of limiting the movement on a 2D plane without gravity, the experiments where conducted on a lubricated glass plate.}
\label{fig:exp}
\end{figure}

The fabrication of the mock-ups is described in \fref{fig:exp}.
% The fabrication of the actuators involved 4 steps: (1) mold casting the main silicone body, opened on one side; (2) wrapping a thin and stiff fiber around the actuator to constraint the radial expansion; (3) incorporating the fabric that acts as inextensible bottom layer by casting an un-cured silicone layer; (4) inserting and securing the nozzles, acting as both air inlets and mechanical clamp.
%
The mold to obtain the main body of the actuator has a semi-cylindrical shape and an insert of the same shape to create the internal chamber (figure \ref{fig:exp} A). Both the main mold and the insert were 3D printed in ABS plastic. We used as silicone rubber the Smooth-on Ecoflex 0030 (tensile modulus at $100\%$ strain $E_{100\%} = 69$ kPa, ultimate strength $\sigma_{Rs} = 1.38$ MPa, elongation at break $900\%$).
We obtained the value of $\mu$ in \tref{tab:mockup} simply as $\mu = (4/3)*E_{100\%}$.
As for the fibre, we used a wire featuring high tensile stiffness and strength (declared ultimate strength $\sigma_{Rw} = 5.8$ GPa), but negligible bending stiffness (diameter $d = 0.10$ mm) to constrain the radial expansion. A fabric with squared stitches made in high-density polyethylene (HDPE), with a tensile modulus $E_f = 1.5$ MPa was used to constraint the axial elongation (figure \ref{fig:exp} B and C).

\section{Mathematical Model}
\label{sec:model}
The goal of this model is to describe both the response of the actuator and its mechanics in terms of stresses and strains, capturing all the nonlinearities deriving from: (1) the hyper-elastic material, (2) the large deformations, (3) the presence of the fibre reinforcements.
We wanted to keep the model easy to read and computationally light. The readability allows the manipulation of the model to adapt it to different structures and provides hints about the physics of the actuator, which can be used to improve its design. As for the computational cost, a light model is suitable for control purposes, a highly required feature in robotics applications.
For all the above reasons, we decided to make simplifying assumptions based on hypotheses on the mechanics of the actuator, which allowed us to obtain a semi-analytical model, where we analytically solve all the differential equations describing the kinematics and the equilibrium of the structure. 
% The only numerical part involved is the solution of high order polynomial equations and integrations of algebraic functions.

The first assumption that we make is to consider the deformation of the structure to follow a toroidal symmetry, as described in figure \ref{fig:toroidal}. In particular, we divide the cross section in three main parts (figure \ref{fig:seg1}): (1) circular, (2) rectangular, (3) corners. We assume the circular part to deform as a portion of torus, while the rectangular part and the boundaries as a portion of cylinder.
This assumption implies to have the curvature of the actuator to be constant in each point along the axis and equal to $\chi = 1/R$, corresponding to the curvature of the surface coincident with the inextensible bottom layer (figure \ref{fig:seg1}).

The model follows the Eulerian description, so we assume a-priori the deformed configuration and we write all the equilibria in this configuration. We then use these equilibria to find the value of the internal pressure that can satisfy them.
As far as the nomenclature is involved, we used a star $.^*$ to indicate quantities in the undeformed configuration that change in the deformed one.

All the assumptions together with their motivations are contained in the remaining of this section, which is organized as follows: we treat individually the three portions and for each of them we first describe and solve the kinematics, then we include the material model and finally we apply the equilibria and solve the equations.

\subsection{Semicircular part}
\subsubsection{Kinematics}
\label{sec:kinc}

%insert a figure:
\begin{figure}[]
\centering
\includegraphics[width=0.8\columnwidth]{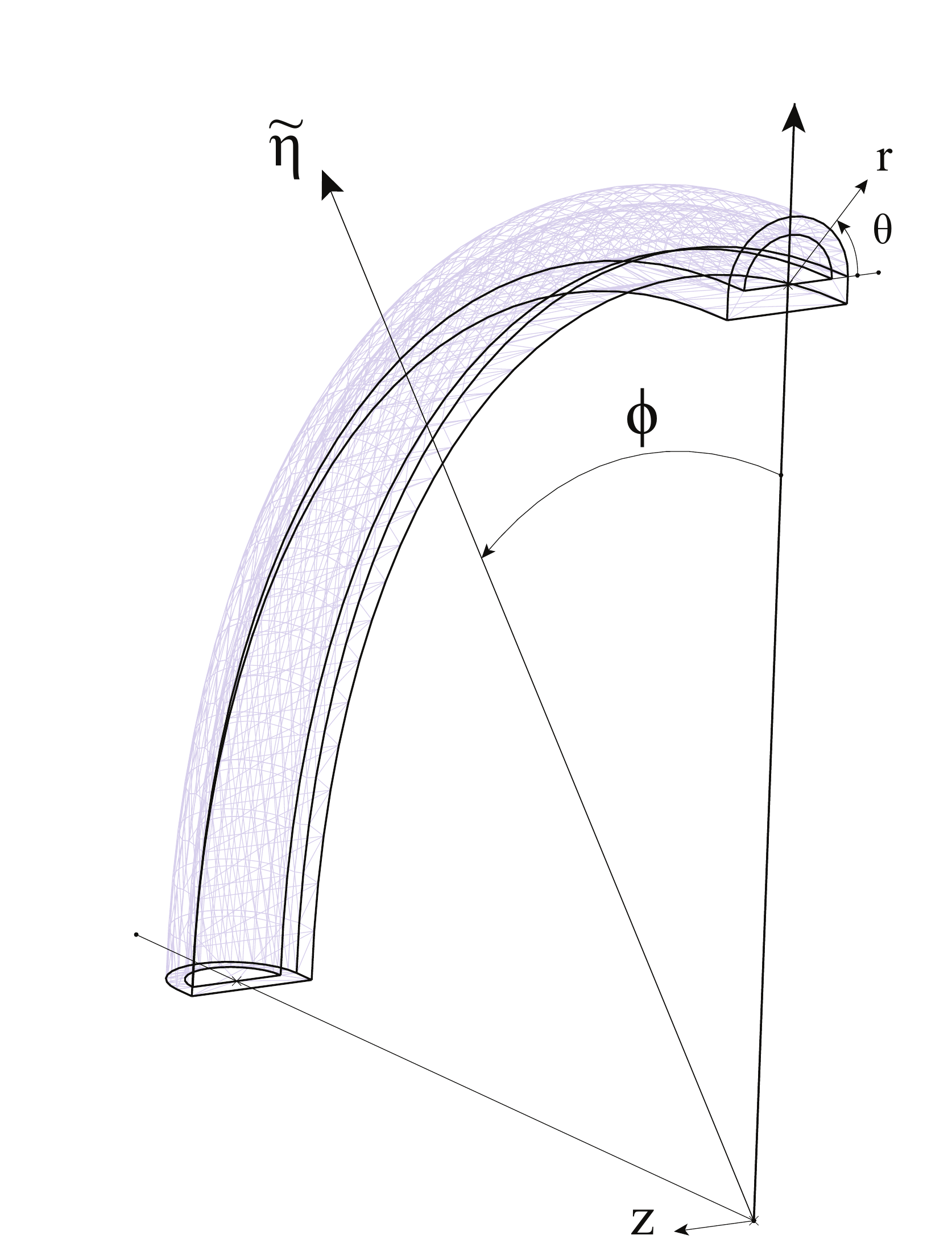}
\caption{Description of the structure deformed following a toroidal symmetry, with the associated toroidal coordinate system $r, \theta, \phi$.}
\label{fig:toroidal}
\end{figure}
We assume the circular portion of the structure (1 in figure \ref{fig:seg1}) to deform into a section of torus defined as (figure \ref{fig:toroidal})
\begin{equation}
R_1 \leq r \leq R_2, \quad 0 \leq \theta \leq \pi, \quad 0 \leq \phi \leq L * \chi.
\end{equation}
In order to define the kinematics of this structure, we reduce the problem to just a half of the section due to symmetry ($0 \leq \theta \leq \pi/2$) and we introduce non-dimensional parameters to describe the geometry:
\begin{equation}
\rho = \frac{r}{R_2} \in \left[k, 1 \right], \quad \theta \in \left[0, \pi/2 \right], \quad \varphi = \frac{\phi}{L * \chi} \in \left[0, 1 \right],
\end{equation}
where $k = R_1 / R_2$.

Because the curvature $\chi$ stays constant along the axis, all the quantities are the same in each cross section ($\frac{\partial}{\partial \varphi} = 0$); on the other hand, the particles deform along $\varphi$ due to the bending. We can describe this deformation by means of $\lambda_\varphi$

Now, in order to describe the kinematics along the $r$ and $\theta$ directions, we need to introduce two functions $g$ and $f$, which map the values of the radius $r$ and angle $\theta$ for a material particle in the deformed configuration back to the values $r^*$ and $\theta^*$ in the undeformed configuration. We will define these functions referred to the non-dimensional parameters
\begin{equation}
g(\rho, \theta) := \rho^*, \quad f(\rho, \theta) := \theta^*.
\end{equation}
Now we can introduce the strains along $r$ and $\theta$, relating them to these functions
%\begin{equation}
%\label{eq:l_r}
%\lambda_\rho = \frac{dr}{dr^*} = \left( \frac{d\rho^*}{d\rho} \right)^{-1} = \left( \frac{d g(\rho, \theta)}{d\rho} \right)^{-1} = \frac{1}{g_\rho(\rho, \theta)}
%\end{equation}
%%
%\begin{equation}
%\label{eq:l_theta}
%\lambda_\theta = \frac{r d\theta}{r^* d\theta^*} = \left( \frac{\rho^*}{\rho} \frac{d\theta^*}{d\theta} \right)^{-1} = \frac{\rho}{g(\rho, \theta)} \frac{1}{f_\theta(\rho, \theta)}
%\end{equation}

\begin{dmath}
\label{eq:l_r_theta}
{\lambda_\rho = \frac{d\rho}{d\rho^*} = \left(\frac{\partial g(\rho,\theta)}{\partial \rho}\right)^{-1}};
\\
{\lambda_\theta = 
\frac{\rho d\theta}{\rho^* d\theta^*} = 
\frac{\rho}{g(\rho, \theta)}
\left(\frac{\partial f(\rho, \theta)}{\partial \theta}\right)^{-1}};
\\
{\lambda_\phi = 1 + \Gamma (\beta^* + \rho \sin \theta )}.
\end{dmath}

Let know apply the incompressibility constraint, obtaining the linear PDE that describes the kinematics of the system
\begin{dmath}
\label{eq:kindiff1}
{\lambda_\varphi \lambda_\rho \lambda_\theta = 1 \Rightarrow 
\rho + \rho\Gamma (\beta^* + \rho \sin \theta )} = g(\rho,\theta) \frac{\partial g(\rho,\theta)}{\partial \rho} \frac{\partial f(\rho, \theta)}{\partial \theta}
\end{dmath}

Considering that $f(\rho, \theta)$ has to fit two boundary conditions on $\theta$:
\begin{equation}
\label{eq:fBC}
f(\rho, 0) = 0, \quad f(\rho, \pi) = \pi
\end{equation}
we assume the displacement of the particles along $\theta$ to be negligible and as a consequence we set $f(\rho, \theta) = \theta$ and thus $\partial f(\rho, \theta) / \partial \theta = 1$.
Substituting in \ref{eq:kindiff1} we obtain
\begin{equation}
\label{eq:kindiff2}
\rho + \rho\Gamma (\beta^* + \rho \sin \theta ) = g(\rho,\theta) \frac{\partial g(\rho,\theta)}{\partial \rho},
\end{equation}
which we can integrate respect to $\rho$ to obtain
\begin{equation}
\label{eq:kin_g_sq}
g^2 = \rho^2 + \rho^2\Gamma\beta^* + \frac{2}{3}\rho^3\Gamma\sin{\theta} + 2c(\theta).
\end{equation}
By applying the boundary condition on $\rho$ ($g(1,\theta) = 1$, meaning that the external surface remains undeformed) we can find $c(\theta)$
\begin{equation}
c(\theta) = -\frac{1}{2}\Gamma\beta^* - \frac{1}{3}\Gamma\sin\theta.
\end{equation}
Finally we find an algebraic expression for $g(\rho,\theta)$
\begin{equation}
\label{eq:kin_g}
g(\rho,\theta) = \sqrt[•]{\rho^2 - \Gamma [\beta^*(1-\rho^2) + (2/3)\sin\theta(1-\rho^3) ]},
\end{equation}
where we excluded the negative case because $g(\rho,\theta) \geq 0$ by definition.

From (\ref{eq:kin_g}) we can see that when the actuator is at rest ($\Gamma = \chi R_2 = 0$), then $g(\rho,\theta) = \rho \:, \forall \theta \in [0, \pi]$.
As far as the conditions of existence for (\ref{eq:kin_g}) are concerned, we need
\begin{equation}
\label{eq:EC}
\rho^2 \geq \Gamma [\beta^*(1-\rho^2) + (2/3)\sin\theta(1-\rho^3)].
\end{equation}

Physically, this condition fixes the minimum limit for the deformed internal radius $R_1$, which is the minimum value of $\rho$, in order to have an undeformed radius $R_1^* \geq 0$. This means that we can make the condition strict.

% Condizioni esistenza
%Now, let set a design constraint, i.e., $\beta^* = b/R_2 \leq 1$. Following it, and considering that $(1-\rho^2)$ and $(1-\rho^3) \geq 0 \forall \rho \in [R_1/R_2, 1]$, the worst condition to respect (\ref{eq:EC}) comes when $\sin \theta = 1$ and $\beta^* = 1$, so we have a new condition, which is more restrictive than (\ref{eq:EC}):
%%
%\begin{equation}
%\label{eq:EC2}
%(2/3)\Gamma \rho^3 + (1+\Gamma) \rho^2 - (5/3)\Gamma > 0
%\end{equation}
%%
%%
%\begin{figure}[]
%\centering
%\includegraphics[scale=1.0]{Figures/cond_rho_2.pdf}
%\caption{Condition of existence for $\rho$ for each value of the design parameter $1$.}
%\label{fig:EC}
%\end{figure}
%%
%Assuming also that the maximum allowed bending angle is $2\pi$ we can obtain a set of three conditions of existence to be respected
%\begin{equation}
%\label{eq:EC_final}
%\begin{matrix}
%\beta^* \geq 1
%\\
%\chi L \leq 2\pi
%\\
%(2/3)\rho^3 + (1+q/(2\pi)) \rho^2 - (5/3) > 0
%\end{matrix}
%\end{equation}
%%
%where $q = L/R_2$ is a design parameter.
%We can obtain the last condition on $\rho$ by associating an equation and solving it numerically. The result is shown in figure \ref{fig:EC}, where for each value of $q$ is shown the minimum value that $\rho$ can assume in order to respect the conditions of existence.

Finally, using (\ref{eq:kin_g_sq}) the value of the lower boundary for $\rho$, i.e., $k = R_1/R_2$ can be computed by solving:
\begin{dmath}
\label{eq:k}
k^3((2/3)\Gamma) + k^2(1+\Gamma\beta^*) - \Gamma(\beta^* + (2/3)\sin\theta) - k^{*2} = 0
\end{dmath}
From (\ref{eq:k}) we can see that: (1) we can compute $k$ knowing the design parameters and the value of $\chi$; (2) $k$ depends on $\theta$, which means that in the deformed configuration, in order to respect the conservation of volume, the arcs formed by the particles in the reference configuration will deform and not stay circular any more;
(3) if $\rho \in [k, 1]$, the condition of existence (\ref{eq:EC}) will be automatically satisfied.
The kinematics of the semi-circular part is now solved, so for each assumed value of $\chi$ we can compute all the strains \eref{eq:l_r_theta} at each point ($\rho, \theta, \phi$).

%Now, we decided to impose another design constraint, which correspond to keep the body slender and it is
%\begin{equation}
%\frac{R_2}{L} \leq \frac{1}{10} \Rightarrow \frac{R_2\chi}{2\pi} = \frac{\Gamma}{2\pi} \leq \frac{1}{10} \Rightarrow \Gamma \leq \frac{2\pi}{10},
%\end{equation}
%where we have also assumed a maximum bending angle $L\chi = 2\pi$.

%
%\begin{equation}
%\lambda_\rho(r, \theta, \chi) = \frac{1}{\frac{\partial \phi}{\partial r}r + \phi}
%\end{equation}
%%
%\begin{equation}
%\lambda_{\theta}(r, \theta, \chi) = \frac{1}{\phi}
%\end{equation}
%%
%\begin{equation}
%\lambda_{\phi}(r, \theta, \chi) = 1 + \chi (b + r \sin\theta)
%\end{equation}
%%
%\begin{equation}
%\phi(r, \theta, \chi) = \frac{r^*}{r}
%= \sqrt[•]{1 + \chi b \left( 1 - \frac{R_2^2}{r^2} \right) + \frac{2}{3}\chi \left( r - \frac{R_2^3}{r^2} \right) \sin \theta},
%\end{equation}
%where $r^*$ is the radius in the undeformed straight configuration, while $r$ is the radius in the deformed one (Eulerian frame).

%From the kinematics, for each value of the curvature $\chi$, we can immediately compute the deformed internal radius $R_1$, which in general varies with $\theta$ as well
%\begin{equation}
%R_1(\theta, \chi) = \frac{R_1^*}{\phi(R_1, \theta, \chi)}
%\end{equation}

\subsubsection{Equilibrium}
The equilibrium in the deformed configuration has to be expressed in toroidal coordinates, coherently with the description of the kinematics.
In this configuration, because the only external action was provided by the internal pressure, we assumed the state of stress in the deformed configuration to be principal, obtaining
%\begin{equation}
%\textbf{T} =
%\begin{pmatrix}
%\tilde{\sigma_{r}} & 0 & 0 \\
%0 & \tilde{\sigma}_{\theta} & 0 \\
%0 & 0 & \tilde{\sigma}_{\phi}
%\end{pmatrix}.
%\end{equation}
\begin{equation}
\textbf{T} =
\left( \begin{array}{ccc}
\tilde{\sigma_{r}} & 0 & 0 \\
0 & \tilde{\sigma}_{\theta} & 0 \\
0 & 0 & \tilde{\sigma}_{\phi}
\end{array} \right)
\end{equation}

We can now apply the Eulerian field equation in static equilibrium and with no body force $\nabla \cdot \textbf{T} = \mathbf{0}$ \cite{ogden1997nonlinear},
%\begin{equation}
%\nabla \cdot \textbf{T} = \mathbf{0},
%\end{equation}
which expressed in toroidal coordinates from \cite{buchanan2005analysis} gives
\begin{equation}
\label{eq:eq1}
\frac{\partial{\tilde{\sigma}_{r}}}{\partial r} + \frac{R + 2r \sin \theta}{r(R + r \sin \theta)} \tilde{\sigma}_{r} 
- \frac{1}{r}\tilde{\sigma}_{\theta} - \frac{\sin \theta}{R+r\sin\theta}\tilde{\sigma}_{\phi} = 0
\end{equation}
\begin{equation}
\label{eq:eq2}
\frac{1}{r}\frac{\partial \tilde{\sigma}_{\theta}}{\partial \theta}
-\frac{\cos\theta}{R+r\sin\theta}\tilde{\sigma}_{\theta}
+\frac{\cos\theta}{R+r\sin\theta}\tilde{\sigma}_{\phi} = 0
\end{equation}
\begin{equation}
\label{eq:eq3}
\frac{\partial\tilde{\sigma}_{\phi}}{\partial\phi} = 0.
\end{equation}
It is interesting to observe that the equilibrium in the radial direction (\ref{eq:eq1}) contains also the stresses $\tilde{\sigma}_{\phi}$ and $\tilde{\sigma}_{\theta}$, as a consequence of the bending of the longitudinal axis from the undeformed cylindrical configuration to the deformed toroidal one. This fact confirms the presence of a coupling between the three principal stresses.

Due to the assumptions that we made on the kinematics, the circumferential (\ref{eq:eq2}) and the axial equilibria \eref{eq:eq3} become redundant, so we will apply only the equilibrium (\ref{eq:eq1}).

We can now nondimensionalize equation (\ref{eq:eq1}) multiplying it by $R_2/\mu$
\begin{equation}
\label{eq:eq1nd}
\frac{\partial{\sigma_\rho}}{\partial \rho} + \frac{1 + 2\Gamma \rho \sin \theta}{\rho(1 + \Gamma \rho \sin \theta)} \sigma_\rho 
- \frac{1}{\rho}\sigma_\theta - \frac{\Gamma \sin \theta}{1+ \Gamma \rho \sin \theta}\sigma_\varphi = 0,
\end{equation}
%%
%\begin{equation}
%\label{eq:eq2nd}
%\frac{\partial \sigma_{\theta}}{\partial \theta}
%-\frac{\Gamma \rho \cos \theta}{1+\Gamma \rho \sin \theta} (\sigma_\phi - \sigma_{\theta})
%= 0
%\end{equation}
%%
where $\sigma_i = \tilde{\sigma}_j/\mu $ are nondimensional stresses.

\subsubsection{Material Model}
\label{sec:mat_c}
As far as the material model is concerned, we wanted to find a trade-off between accurateness in the description of the behaviour and compactness, to allow easy manipulation in an analytical model. We discarded a linear material model because one of the key points of this work is to capture the nonlinear response of the actuators. We considered then the Neo-hookean material model \cite{ogden1997nonlinear} and the Gent model \cite{Gent1996}. The first one has a relatively compact formulation and can replicate accurately the initial response of hyper-elastic rubbers, while it totally fails in capturing the stiffening that happens when the molecules become almost totally stretched. The Gent model, on the other hand, can accurately capture all the nonlinear response of rubbers but it has a slightly more complicated formulation. By comparing the two models, we could observe that the discrepancy in the behaviour happens for $\lambda > 3$. From a first estimation we expected the strains to be $\lambda \leq 3$ for our actuators in the chosen ranges of design parameters, so we decided to use the Neo-Hookean model and then to verify a-posteriori the goodness of our choice.

The constitutive equations for the incompressible Neo-Hookean materials are then
%\begin{equation}
%\label{eq:mat1}
%\sigma_r(\rho, \theta, \chi, p) = \lambda_\rho - \frac{Q}{\lambda_\rho}
%\end{equation}
%%
%\begin{equation}
%\label{eq:mat2}
%\sigma_\theta(\rho, \theta, \chi, p) = \lambda_\theta - \frac{Q}{\lambda_\theta}
%\end{equation}
%%
%\begin{equation}
%\label{eq:mat3}
%\sigma_\varphi(\rho, \theta, \chi, p) = \lambda_\varphi - \frac{Q}{\lambda_\varphi}.
%\end{equation}
%%
%
\begin{equation}
\label{eq:mat}
\sigma_i(\rho, \theta, \chi, p) = \lambda_i - Q/\lambda_i,
\quad
\textrm{with}
\:
i = \{ \rho, \theta, \varphi \}
\end{equation}
where $Q(\rho, \theta, \chi, p) = \tilde{Q}/\mu$ is a nondimensional lagrangian parameter.

\subsubsection{Solution}
We can rewrite equation (\ref{eq:eq1nd}) as
\begin{equation}
\label{eq:eq1nd2}
\frac{\partial{\sigma_\rho}}{\partial \rho} + a_1 \sigma_\rho 
- \frac{1}{\rho}\sigma_\theta - a_2 \sigma_\varphi = 0
\end{equation}
%
%\begin{equation}
%\label{eq:eq2nd2}
%\frac{\partial \sigma_{\theta}}{\partial \theta}
%-a_3 (\sigma_\phi - \sigma_{\theta})
%= 0
%\end{equation}

with
%\begin{equation}
%\begin{matrix}
%a_1(\rho,\theta) =& \frac{1 + 2\Gamma \rho \sin \theta}{\rho(1 + \Gamma \rho \sin \theta)}
%\\
%\\
%a_2(\rho,\theta) =& \frac{\Gamma \sin \theta}{1+ \Gamma \rho \sin \theta}
%\end{matrix}
%\end{equation}
%
\numparts \begin{eqnarray}
a_1(\rho,\theta) =& 1 + 2\Gamma \rho \sin \theta/(\rho(1 + \Gamma \rho \sin \theta))
\\
a_2(\rho,\theta) =& \Gamma \sin \theta/(1+ \Gamma \rho \sin \theta)
\end{eqnarray} \endnumparts

%\begin{equation}
%\begin{matrix}
%a_1(\rho,\theta) =& \frac{1 + 2\Gamma \rho \sin \theta}{\rho(1 + \Gamma \rho \sin \theta)}
%\\
%\\
%a_2(\rho,\theta) =& \frac{\Gamma \sin \theta}{1+ \Gamma \rho \sin \theta}
%\\
%\\
%a_3(\rho,\theta) =& \frac{\Gamma \rho \cos \theta}{1+\Gamma \rho\sin \theta}
%\end{matrix}
%\end{equation}

Substituting (\ref{eq:mat}) into (\ref{eq:eq1nd2}), we obtain
\begin{equation}
\label{eq:Q}
\frac{\partial Q}{\partial \rho} =
Q (b_2/b_1) + b_3/b_1
\end{equation}
with
%\begin{equation}
%\begin{matrix}
%b_1(\rho,\theta) =&  -\frac{1}{\lambda_\rho} 
%\\
%\\
%b_2(\rho,\theta) =& \frac{1}{\lambda_\rho^2}\frac{\partial \lambda_\rho}{\partial \rho} - \frac{a_1}{\lambda_\rho} + \frac{1}{\rho \lambda_\theta} + \frac{a_2}{\lambda_\phi} 
%\\
%\\
%b_3(\rho,\theta) =& \frac{\partial \lambda_\rho}{\partial \rho} + a_1\lambda_\rho - \frac{\lambda_\theta}{\rho} - a_2\lambda_\phi
%\end{matrix}
%\end{equation}
%
\numparts \begin{eqnarray}
b_1(\rho,\theta) =&  -1 /\lambda_\rho
\\
b_2(\rho,\theta) =& \frac{1}{\lambda_\rho^2}\frac{\partial \lambda_\rho}{\partial \rho} - \frac{a_1}{\lambda_\rho} + \frac{1}{\rho \lambda_\theta} + \frac{a_2}{\lambda_\varphi} 
\\
b_3(\rho,\theta) =& \frac{\partial \lambda_\rho}{\partial \rho} + a_1\lambda_\rho - \frac{\lambda_\theta}{\rho} - a_2\lambda_\varphi
\end{eqnarray} \endnumparts

Equation \eref{eq:Q} is a linear first order differential equation, which solution is
\begin{equation}
Q(\rho, \theta) = \frac{1}{\tau(\rho,\theta)} \left[ Q(k, \theta) + \int_{k}^\rho \tau(\tilde{\rho}, \theta) \frac{b_3(\tilde{\rho},\theta)}{b_1(\tilde{\rho}, \theta)} d\tilde{\rho} \right]
\end{equation}
with
\begin{equation}
\tau(\rho, \theta) = \exp \left( \int_{k}^\rho \frac{b_2(\tilde{\rho},\theta)}{b_1(\tilde{\rho},\theta)} d\tilde{\rho} \right).
\end{equation}

We need now a boundary condition on the radial stress $\sigma_\rho$, which we can apply at the inner chamber where it has to equal the internal pressure
\begin{equation}
\sigma_\rho(k, \theta) = -p/\mu \quad \forall \theta \in [0, \pi/2].
\label{eq:bcr}
\end{equation}
Substituting equation (\ref{eq:mat}) into the boundary condition (\ref{eq:bcr}), we get
\begin{equation}
Q(k, \theta) = (p/\mu) \lambda_\rho(k, \theta) + \lambda_\rho^2(k, \theta).
\end{equation}
The problem for the semi-circular part of the cross section is now closed: knowing $\chi$ and the strains obtained from \eref{eq:l_r_theta}, we can compute all the stresses \eref{eq:mat} for a given value of the internal pressure $p$.

\subsection{Rectangular part}

\subsubsection{Kinematics}
\label{sec:kinr}
The rectangular part represents the wall between the chamber and the bottom layer (2 in figure \ref{fig:seg1}). We assume that this portion of the section, whose domain is defined in \eref{eq:domain_r}, deforms as a consequence of the internal pressure, while the corners at the boundaries (3 in figure \ref{fig:seg1}) remain unchanged. We will treat the kinematics of the corners in section \ref{sec:corner}.
The rectangular portion when deformed becomes a sector of a hollow cylinder defined as (figure \ref{fig:toroidal})
\begin{equation}
\label{eq:domain_r}
-R_1 \leq z \leq R_1, \quad 1/\chi \leq \tilde{\eta} \leq h + 1/\chi, \quad 0 \leq \phi \leq \chi L.
\end{equation}
Non-dimensionalizing the variables and considering only half a cylinder due to symmetry we get
\begin{dmath}
{\zeta = z/R_2 \in [0, k(0)], \quad \eta = (\tilde{\eta} - 1/\chi)/R_2 \in [0, \beta],}
\\
\varphi = \frac{\phi}{L * \chi} \in \left[0, 1 \right].
\end{dmath}

In this domain, we can define the strains. The longitudinal strain is
\begin{equation}
\label{eq:lphi_r}
\lambda_\varphi = 1 + \Gamma\eta.
\end{equation}
For the transverse strain, due to the constraint represented by the external fibres on both sides of the structure, we assume there is no movement of particles in the $\zeta$ direction and so
$\lambda_\zeta = 1$.
Now, we can obtain $\lambda_\eta$ from the conservation of volume
\begin{equation}
\label{eq:kin_r}
\lambda_\varphi \lambda_\zeta \lambda_\eta = 1 \Rightarrow \lambda_\eta = 1/\lambda_\varphi = 1/(1+\Gamma\eta).
\end{equation}
In order to find the deformed thickness $\beta$ we can integrate $\lambda_\eta$
\begin{dmath}
{\lambda_\eta = d\eta / d \eta^* \Rightarrow\int_0^{\beta} d\eta/\lambda_\eta =}
\\
\int_0^{\beta^*} d\eta^*
\Rightarrow  \beta + \Gamma \beta^2 /2 = \beta^*
\end{dmath}
Solving we get
%\begin{dmath}
%\beta_{1/2} = -1/\Gamma \pm \sqrt[•]{(1/\Gamma^2) + (2\beta^*/\Gamma)} \Rightarrow
%\beta = (\sqrt[•]{1+2\Gamma\beta^*} - 1)/\Gamma
%\end{dmath}
\begin{dmath}
\beta = (\sqrt[•]{1+2\Gamma\beta^*} - 1)/\Gamma
\end{dmath}
were we select only the positive solution because we want $\beta \geq 0$.
With this result we completed the kinematic analysis and we can obtain all the strains in the rectangular part \eref{eq:kin_r} for each given value of $\chi$.

\subsubsection{Equilibrium}
The only non-trivial equilibrium equation expressed in cylindrical coordinates is
\begin{equation}
\frac{ d \tilde{\sigma}_{\tilde{\eta}}}{d \tilde{\eta}} + \frac{\tilde{\sigma}_{\tilde{\eta}} - \tilde{\sigma}_\phi }{\tilde{\eta}} = 0,
\end{equation}
which transformed in non-dimensional form becomes
\begin{equation}
\label{eq:eq_r}
\frac{d \sigma_\eta}{d \eta} + \frac{\sigma_\eta - \sigma_\varphi}{\eta + 1/\Gamma} = 0.
\end{equation}

\subsubsection{Material Model}
Assuming identical constitutive model as in section \ref{sec:mat_c} we obtain
\begin{dmath}
\label{eq:matr}
{
\sigma_\eta = \lambda_\eta - \frac{G}{\lambda_\eta},
}
\:
{
\sigma_\zeta = 1 - G,
}
\:
\\
{
\sigma_\varphi = \lambda_\varphi - \frac{G}{\lambda_\varphi} = \frac{1}{\lambda_\eta} - G\lambda_\eta
}
\end{dmath}

where $G(\eta, \chi, p) = \tilde{G}/\mu$ is a nondimensional lagrangian parameter.

\subsubsection{Solution}
Rewriting (\ref{eq:eq_r}) as 
\begin{equation}
d \sigma_\eta /d \eta  + \Gamma\lambda_\eta(\sigma_\eta - \sigma_\varphi) = 0
\end{equation}
and using the constitutive equations (\ref{eq:matr}), we obtain
\begin{equation}
\frac{d \lambda_\eta}{d \eta} - \frac{1}{\lambda_\eta}\frac{d G}{d \eta} + \frac{G}{\lambda_\eta^2}\frac{d \lambda_\eta}{d \eta} + \Gamma(\lambda_\eta^2 - 1) + G \Gamma(\lambda_\eta^2 - 1).
\end{equation}
Including the kinematic expression (\ref{eq:kin_r}) differentiated respect to $\eta$
\begin{equation}
d \lambda_\eta / d \eta = - \Gamma \lambda_\eta^2,
\end{equation}
and rearranging we obtain
\begin{equation}
\label{eq:diff_rett}
d G / d \eta = G \Gamma(\lambda_\eta^2 - 2) - \Gamma,
\end{equation}
a linear first order ODE, which solution requires one boundary condition on $\sigma_\eta$, i.e.,
\begin{equation}
\label{eq:diff_rett_bc}
\sigma_\eta(\beta) = -p/\mu \Rightarrow G(\beta) = \lambda_\eta^2(\beta) + (p/\mu)*\lambda_\eta(\beta).
\end{equation}
The problem in (\ref{eq:diff_rett}) and (\ref{eq:diff_rett_bc}) is a final value problem, so in order to integrate it, let transform it to an initial value problem by setting
\begin{dmath}
{
\bar{\eta} = - \eta \in [-\beta, 0];
\:
\bar{G}(\bar{\eta}) = G(\eta);
\:
d \bar{G} / d \bar{\eta}
}
= - d G / d \eta;
\:
\lambda_{\bar{\eta}} = 1/(1-\Gamma\bar{\eta}). 
\end{dmath}
The new initial value problem is now represented by
%\begin{equation}
%\label{eq:diff_ivp}
%\begin{matrix}
%d \bar{G} / d \bar{\eta} = - \bar{G} \Gamma(\lambda_{\bar{\eta}}^2 - 2) + \Gamma
%\\
%\bar{G}(-\beta) = \lambda_{\bar{\eta}}^2(-\beta) + (p/\mu)*{\bar{\eta}}(-\beta) = (\mu+p(1+\Gamma\beta))/(\mu(1+\Gamma\beta)^2),
%\end{matrix}
%\end{equation}
%
\begin{dmath}
\label{eq:diff_ivp}
d \bar{G} / d \bar{\eta} = - \bar{G} \Gamma(\lambda_{\bar{\eta}}^2 - 2) + \Gamma,
\\
\bar{G}(-\beta) = \lambda_{\bar{\eta}}^2(-\beta) + (p/\mu)*{\bar{\eta}}(-\beta) = (\mu+p(1+\Gamma\beta))/(\mu(1+\Gamma\beta)^2),
\end{dmath} 

and its solution is
\begin{equation}
\label{eq:diff_ivp_sol}
\bar{G}(\bar{\eta}) = (1/\tau(\bar{\eta})) [\bar{G}(-\beta) + \int_{-\beta}^{\bar{\eta}} \Gamma * \tau(\bar{\eta}) \: d \bar{\eta}] = G(\eta)
\end{equation}
with
\begin{dmath}
\label{eq:tau_bar}
\tau(\bar{\eta}) = \exp \left( \int_{-\beta}^{\bar{\eta}} - \Gamma (\lambda_{\bar{\eta}}^2 - 2) d\bar{\eta} \right) = \exp\left( 2\Gamma(\beta + \bar{\eta}) + \frac{1}{1+\Gamma \beta} - \frac{1}{1-\Gamma\bar{\eta}} \right)
\end{dmath}

By substituting the boundary condition $\bar{G}(-\beta)$ from (\ref{eq:diff_ivp}) and $\tau(\bar{\eta})$ from (\ref{eq:tau_bar}), the integral in (\ref{eq:diff_ivp_sol}) has to be computed numerically, giving the values of $G(\eta) = \bar{G}(\bar{\eta})$. We have now completed the analysis of the rectangular part, so we can compute all the strains \eref{eq:kin_r} and stresses \eref{eq:matr} for given values of $\chi$ and $p$.

\subsection{Corners}
\label{sec:corner}
For a matter of simplicity, we restrict the kinematics of the corners (3 in figure \ref{fig:seg1}). In particular, we assume that for the corners the conservation of the volume does not hold. This statement may seem physically not plausible, but we should consider that both the circular part and the rectangular part will in fact move some of their particles into the corners. 
But, if we followed these particles, the kinematic description for both the circular part and the rectangular part would be much more complicated than what we obtained in the previous sections. In order to keep an acceptable level of accuracy and at the same time to obtain an elegant mathematical model, we then decided to simplify the description of the corners.

With these assumptions, each of the corners deforms as a portion of hollow cylinder. We consider only one for symmetry, which description is
\begin{dmath}
{
R_1 \leq z \leq R_2,
1/\chi \leq \tilde{\eta} \leq h^* + 1/\chi,
}
\:
0 \leq \phi \leq \chi L.
\end{dmath}

Non-dimensionalizing the variables we get
\begin{dmath}
{
\zeta = z/R_2 \in [k(0), 1];
\:
\eta = (\tilde{\eta} - 1/\chi)/R_2 \in [0, \beta^*]; 
}
\\
{
\varphi = \phi / (L * \chi) \in \left[0, 1 \right].
}
\end{dmath}

Following the previous assumptions we obtain
$
\lambda_\zeta = 1, \quad \lambda_\eta = 1.
$
The only non-zero deformation is then
\begin{equation}
\lambda_\varphi = 1 + \Gamma\eta.
\end{equation}

For the stress in the corners we then simply have
\begin{equation}
\sigma_\varphi = \lambda_\varphi - 1/\lambda_\varphi = 1 + \Gamma\eta - 1/(1+\Gamma\eta).
\end{equation}

\subsection{Rotational equilibrium}
In a first model for BFAs, Polygerinos et al. \cite{polygerinos2015modeling} reduced all the effects of the internal pressure $p$ to a concentrated torque $M_t$ acting at the tip of the actuator, generated as the integration of the infinitesimal torques on the transverse section at the tip respect to the inextensible bottom layer:
\begin{equation*}
\tilde{M}_t(p) 
= - 2p \int_0^{\pi/2}\int_0^{R_1} (r \sin \theta + b) r\, dr \, d\theta,
\end{equation*}
%= - \frac{p}{6} (3\pi R_1^2 b + 4 R_1^3) \; .$
where in our  case $R_1 = R_1(\theta)$.
From this equation, we can notice that $M_t$ depends linearly on $p$ but does not depend on the curvature $\chi$. While the torque $M_t$ can be considered the main action that produces the bending when the actuator is at rest ($\chi=0$), it fails to capture its behaviour when the deformation increases. In facts, from experimental results we obtained that these actuators show a highly nonlinear response for $0 \leq \chi L \leq 3\pi/2$ (the typical range of application) as shown in figure \ref{fig:validation}.
One essential element lacking to reproduce this trend is the pressure force at the tip $\tilde{F}_t = 0.5p\pi R_1^2$, which generates a moment $M_{\chi}$ respect to each section $\varphi$. $M_{\chi}$, non-zero for $\chi > 0$ and depending on $\chi$, respect to a generic section $\varphi$ equals (see figure \ref{fig:seg1})
\begin{equation}
\label{eq:Mtil_chi_phi}
\tilde{M}_{\chi\varphi}(p,\varphi, \chi) = \tilde{F}_t(p) L \frac{\sin\left( (\chi L/2)* \left(1-\varphi \right) \right)^2}{\chi L/2} .
\end{equation}
%For the scope of our work, we assume the torque to be constant in each section $\phi$ and equal to
%\begin{equation}
%\tilde{M}_{\chi} = \frac{2 F_t(p)}{\chi} \int_0^1 \frac{\sin\left( \frac{\chi L}{2} \left(1-\varphi \right) \right)^2}{(1-\varphi)} d\varphi = \frac{2 F_t(p)}{\chi} \int_0^{\chi L /2} \frac{\sin(t)}{t} d t
%\end{equation}
%where we used mean value theorem for definite integrals.

%at the base equals
%\begin{equation}
%M_b(p,\varphi) = 2 F_t(p) L \frac{\sin(\varphi/2)^2}{\varphi}
%\end{equation}

%%
%\begin{figure}[tb]
%\centering
%\includegraphics[scale=0.35]{Figures/statica.pdf}
%\caption{Bending angle $\phi$ for different values of the applied pressure in quasi-statics conditions. Comparison between the theoretical and experimental results.}
%\label{fig:statics}
%\end{figure}
%%

Summarizing, the pressure at the tip of the actuator produces a force $F_t$, directed as the tangent at the tip and applied at the centroid of the transverse section of the chamber; however, because the neutral surface is forced to coincide with the inextensible layer, the transport of $F_t$ to the neutral surface produces also the moment $M_t$.

Now, the non-dimensional internal elastic torque, constant in each of the sections $\varphi$, is
%\begin{equation}
%\begin{matrix}
%M_\sigma(\chi) 
%= 2 \int_0^{\pi/2} \int_{k(\theta)}^{1} \sigma_{\varphi}^c (\rho \sin \theta + \beta^*) \rho \, d\rho \, d\theta +
%\\
%2 \int_0^{k(0)}\int_0^{\beta} \sigma_{\varphi}^r * \eta \: d \eta d \zeta + 
%2 \int_{k(0)}^{1}\int_0^{\beta^*} \sigma_{\varphi}^b * \eta \: d \eta d \zeta.
%\end{matrix}
%\end{equation}
%
\begin{dmath}
M_\sigma(\chi) 
= 2 \int_0^{\pi/2} \int_{k(\theta)}^{1} \sigma_{\varphi}^c (\rho \sin \theta + \beta^*) \rho \, d\rho \, d\theta +
\\
2 \int_0^{k(0)}\int_0^{\beta} \sigma_{\varphi}^r * \eta \: d \eta d \zeta + 
2 \int_{k(0)}^{1}\int_0^{\beta^*} \sigma_{\varphi}^b * \eta \: d \eta d \zeta.
\end{dmath}
where apexes $c$, $r$, and $b$ refer to circular part, rectangular part and boundaries (corners) respectively.

Let now non-dimensionalize and compute the external torques. For the transport torque $\tilde{M_t}$, first of all we need to add the contribution of the area created by the compression of the rectangular part
\begin{dmath}
\tilde{M}_t
= - 2p \int_0^{\frac{\pi}{2}}\int_0^{R_1} (r \sin \theta + b) r\, dr \, d\theta - p R_1(0)*(h^*-h)^2
\end{dmath}
and then we can non-dimensionalize dividing by $\mu R_2^3$ and integrate along $\rho$
%
%\begin{equation}
%\begin{matrix}
%M_t(p) = \tilde{M}_t(p) / (\mu R_2^3) = -2(p/\mu)\int_0^{\pi/2} \int_{k(\theta)}^1 (\rho \sin\theta + \beta^*)\rho d \rho d \theta 
%-(p/\mu) k(0) (\beta^*-\beta)^2
%= 
%\\
%= -2(p/\mu)\int_0^{\pi/2} \left[\frac{(1 - k(\theta)^3)}{3}\sin\theta + \frac{(1 - k(\theta)^2)}{2}\beta^* \right] d\theta
%-(p/\mu) k(0) (\beta^*-\beta)^2
%\end{matrix}
%\end{equation}
%
\begin{dmath}
M_t(p) 
= -2\frac{p}{\mu}\int_0^{\pi/2} \left[\frac{(1 - k(\theta)^3)}{3}\sin\theta + \frac{(1 - k(\theta)^2)}{2}\beta^* \right] d\theta
-(p/\mu) k(0) (\beta^*-\beta)^2,
\end{dmath}
%
%\begin{equation}
%M_t(p) = \tilde{M}_t(p) / (\mu R_2^3) 
%= - \frac{p}{6\mu} (3\pi k^2 \beta^* + 4 k^3) \;
%\end{equation}
where the last integral has to be computed numerically due to the presence of $k(\theta)$.

Non-dimensionalizing the moment of the force $F_t$ (\ref{eq:Mtil_chi_phi}), we have instead
\begin{dmath}
M_{\chi\varphi}(p,\varphi,\Gamma) = q\frac{\tilde{F}_t(p)}{\mu R_2^2} \frac{\sin\left( \frac{\Gamma q}{2} \left(1-\varphi \right) \right)^2}{\Gamma q/2} \; .
\end{dmath}
For the scope of our work we assumed the actuator to deform with constant curvature, so all the torques will be the same in each section, but in reality $M_{\chi\varphi}$ depends on $\varphi$.
For this reason, we decided to apply all the equilibria in the middle section, in which $\varphi=0.5$, so
\begin{equation}
M_{\chi} = M_{\chi\varphi}(p,0.5,\Gamma) = - F_t \frac{q}{2} \frac{\sin\left( \Gamma q /4 \right)^2}{\Gamma q/2}
\end{equation}
%\begin{equation}
%M_{\chi} = M_{\chi\varphi}(p,0.5,\Gamma) = - F_t (q/2) \sin\left( \Gamma q /4 \right)^2/(\Gamma q/2)
%\end{equation}
%
with
\begin{dmath}
F_t = \tilde{F}_t /(\mu R_2^2) = 2(p/\mu) \int_0^{\pi/2} \int_{k(\theta)}^1 \rho d\rho d\theta + (p/\mu) k(0) (\beta-\beta^*) 
= (p/\mu) \int_0^{\pi/2} (1 - k(\theta)^2) d\theta + (p/\mu) k(0) (\beta-\beta^*)
\end{dmath}
where again the last integral with $k(\theta)$ has to be computed numerically.

At the end, our equilibrium equation becomes

\begin{equation}
\label{eq:equilibrium}
M_\sigma + M_t + M_{\chi} = 0
\end{equation}

\subsection{Solution}
We have now all the equations to describe the kinematics and statics of our actuator. Interestingly, we could solve analytically all the differential equations involved in the model, while we need to use numerical computation to evaluate some algebraic integration.
The procedure that we decided to follow to solve the equations is the following.
We set a value of the curvature $\chi=\Phi/L$, being $\Phi$ the angle between the tangent at the base and the tangent at the tip of the actuator (figure \ref{fig:seg1}).
The range of $\Phi$ that spans all the physically possible configurations of the actuator when bending in free state is $[0, 3\pi/2]$.
For each $\chi$, we can compute all the kinematic quantities including strains and displacements.
In order to compute the quantities that vary on the plane of the cross section, we created three equally spaced grids of points for $\rho$ and $\theta$ in the circular section and $\eta$ and $\zeta$ in the rectangular section and in the corners (figure \ref{fig:disp}).

Next, we can obtain the solution of the statics by starting with a guessed value of the internal pressure, called $p_g$. Using this value and the already obtained kinematics, we can compute all the stresses on the section using the same grids defined for the kinematics and finally we can compute the error function for the equilibrium, definite as
$F(\chi,p_g) = M_\sigma + M_t + M_{\chi}$.
%\begin{equation}
%F(\chi,p_g) = M_\sigma + M_t + M_{\chi}.
%\end{equation}
Using an iterative algorithm we can now solve the problem of finding the value of $p$ that guarantees the equilibrium, for each $\chi$, by setting
\begin{equation}
\label{eq:iterative}
F(\chi,p) = M_\sigma + M_t + M_{\chi} = 0.
\end{equation}

The model and its solution are now closed. In the next sections we will describe the experimental set-up (section \ref{sec:exp}) used to validate the model (section \ref{sec:validation}) and the results on the mechanics of the structure (section \ref{sec:stress_strain}) and on its characterization as a soft actuator (section \ref{sec:characterization}).

%We decide a value for the curvature $\chi$, which we compute the kinematics in closed form ($g$ and strains), and then we use Newton-Rapson to obtain the value of pressure $p$ which corresponds to zero of the F function defined as
%\begin{equation}
%F(\chi,p) = M_\sigma + M_t + M_{\chi}
%\end{equation}
%starting with a guessed value $p_g$.

\section{Experimental set-up}
\label{sec:exp}
The experimental set-up (Fig. 6) was composed by: 
(1) one proportional electronic valve (Camozzi K8P), Input: $0 - 10$ V, Output: $0 - 3$ bar; (2) an Arduino MEGA microcontroller; (3) a camera for data acquisition; (4) the mock-up.
\begin{figure}[]
\centering
\includegraphics[scale=1.0]{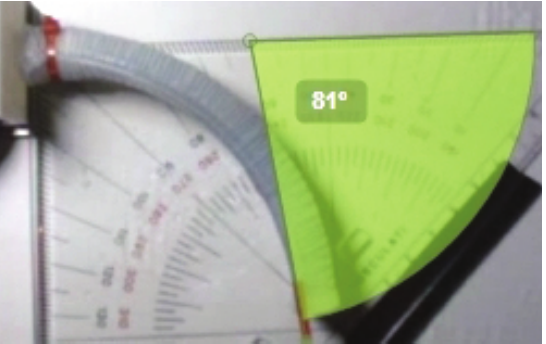}
\caption{Actuator in bended state with angle estimated using Kinovea software for image analysis.}
\label{fig:bend}
\end{figure}
Each experiment was conducted starting from rest position of the actuator and increasing the pressure up to the desired value through the electronic signal provided by the Arduino microcontroller to the electronic valve. The actuator bended as a consequence of the increase in internal pressure and its equilibrium configuration was recorded using the camera. From the pictures we computed the bending angle using Kinovea software for image analysis (figure \ref{fig:bend}), with a resolution of $1^\circ$.
Each experiment was repeated 4 times and we computed the average and the standard deviation. 

\section{Results}
\label{sec:results}
\subsection{Numerical solution}
We solved the iterative problem (\ref{eq:iterative}) using MATLAB and the Trust-Region-Reflective Algorithm \cite{More1983}, \cite{Steihaug1983}.

A big advantage of this model on the side of computational time is that all the kinematics can be solved once for all for each values of $\chi$ and the results can be stored in a database.
As for the statics, using 610 points on the grids of the cross-section, MATLAB and a simple algorithm (not optimized for fast computation) running on a 2.9 GHz Intel Core i5 computer, each step of the statics is obtained in $\approx0.05$ s.

\subsection{Validation}
\label{sec:validation}
%%
%\begin{figure}[]
%\centering
%\includegraphics[scale=0.6]{Figures/valid_1_pl.pdf}
%\caption{Validation of the model on the experimental data of the first mock-up.}
%\label{fig:valid1}
%\end{figure}
%%
%%
%\begin{figure}[]
%\centering
%\includegraphics[scale=0.6]{Figures/valid_2_pl.pdf}
%\caption{Validation of the model on the experimental data of the second mock-up.}
%\label{fig:valid2}
%\end{figure}
%%
%%
%\begin{figure}[]
%\centering
%\includegraphics[scale=0.6]{Figures/valid_3_pl.pdf}
%\caption{Validation of the model on the experimental data of the third mock-up.}
%\label{fig:valid3}
%\end{figure}
%
%%
%\begin{figure}[]
%\centering
%\includegraphics[scale=0.9]{Figures/validation.pdf}
%\caption{Validation of the model on the experimental data for the first (a), second (b) and third (c) mock-ups (\tref{tab:mockup}). For the experimental values we plotted the average and standard deviation as circles and errorbars. We computed the full model as well as the model with the pressure on the lateral surface of the chamber artificially switched off.}
%\label{fig:validation}
%\end{figure}
%%
%
\begin{figure*}[]
\centering
\includegraphics[scale=1.0]{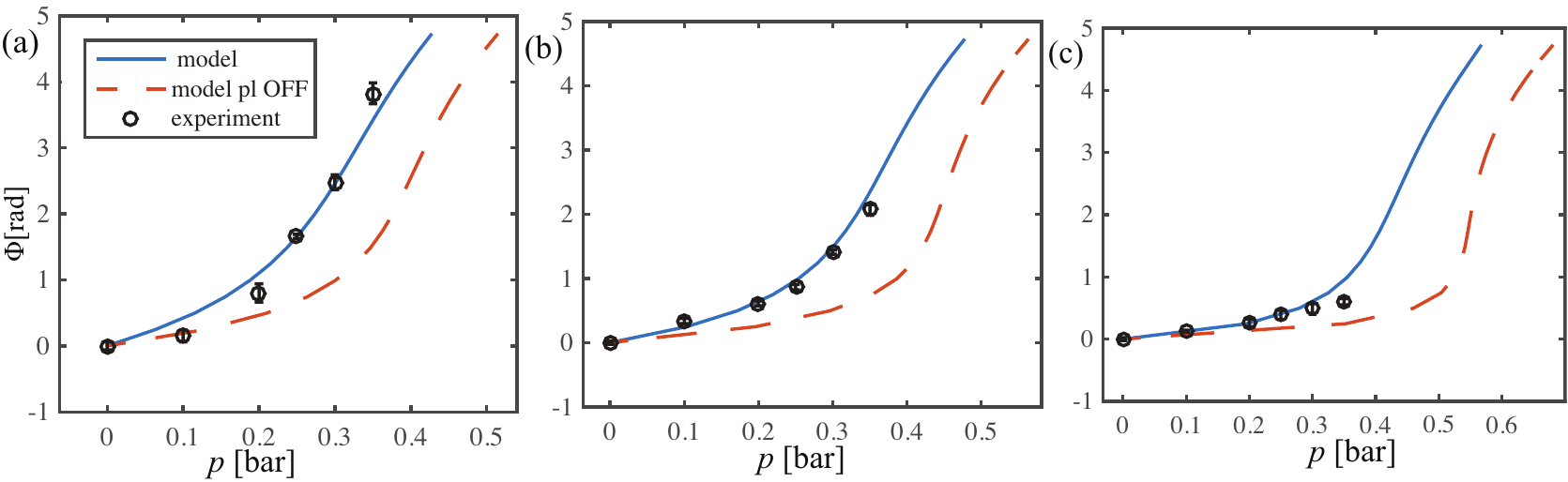}
\caption{Validation of the model on the experimental data for the first (a), second (b) and third (c) mock-ups (\tref{tab:mockup}). For the experimental values we plotted the average and standard deviation as circles and errorbars. We computed the full model as well as the model with the pressure on the lateral surface of the chamber artificially switched off.}
\label{fig:validation}
\end{figure*}
In order to validate the model we compared its results with the experimental ones. As for the model, we span the $\Phi$ angle in $[0, \frac{3}{2}\pi]$ and for each value we obtained the pressure $p$ that guarantees the equilibrium of the structure.
The experimental results were instead obtained by driving the actuator with a pressure p in $[0, 0.35]$ bar and acquiring the correspondent angle $\Phi$ at the equilibrium.
The validation of the model on the three mock-ups is shown in figure \ref{fig:validation}.
It is possible to observe that the behaviour of the actuator changes considerably between the three mock-ups and that the model is able to capture the response in all of the cases.
It should be noticed that the validation is successful using the nominal parameters for the geometry and the material, without any fitting procedure, confirming the physical plausibility of our assumptions.

We wanted also to explore the importance of the pressure on the lateral surface of the chamber respect to the one on the tip. We then artificially set to zero the pressure on the lateral surface and simulated the model in the same interval. The results can be found in comparison with the full model and the experiments in figure \ref{fig:validation} under the name ``model pl OFF".

\subsection{Strains and stresses}
\label{sec:stress_strain}
\begin{figure}[]
\centering
\includegraphics[width=0.9\columnwidth]{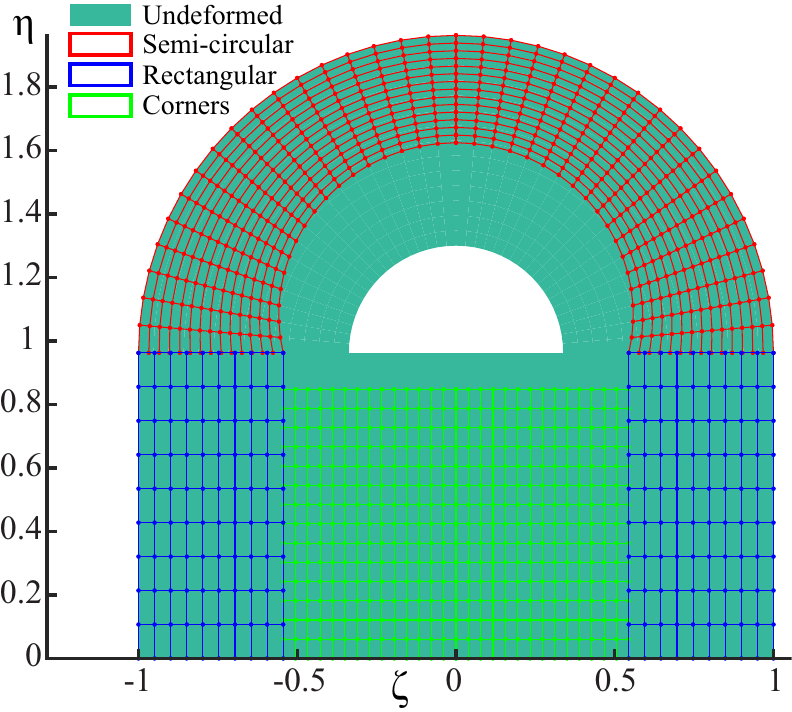}
\caption{Deformed state of the cross section with the equally-spaced grids used for computation, superimposed to the undeformed one.}
\label{fig:disp}
\end{figure}

In this section we analyzed the state of deformation and stress on the cross sections of the structure.
We used the same parameters of the mock-up number 2 (table \ref{tab:mockup}) for a bending angle $\Phi = 3/2 \pi$.
In figure \ref{fig:disp} we can visualize the deformed cross section, with the equally-spaced grid used for the numerical computation, overlapped over the undeformed cross section. The deformation of the chamber is clearly macroscopic and follows the assumptions made about the kinematics in sections \ref{sec:kinc}, \ref{sec:kinr} and \ref{sec:corner}.
As assumed, we can observe a sliding between the rectangular part, the corners and the circular part, which we used as a simplification of the behaviour of the boundaries that connect the three different portions of the cross section.
\begin{figure}[]
\centering
\includegraphics[width=\columnwidth]{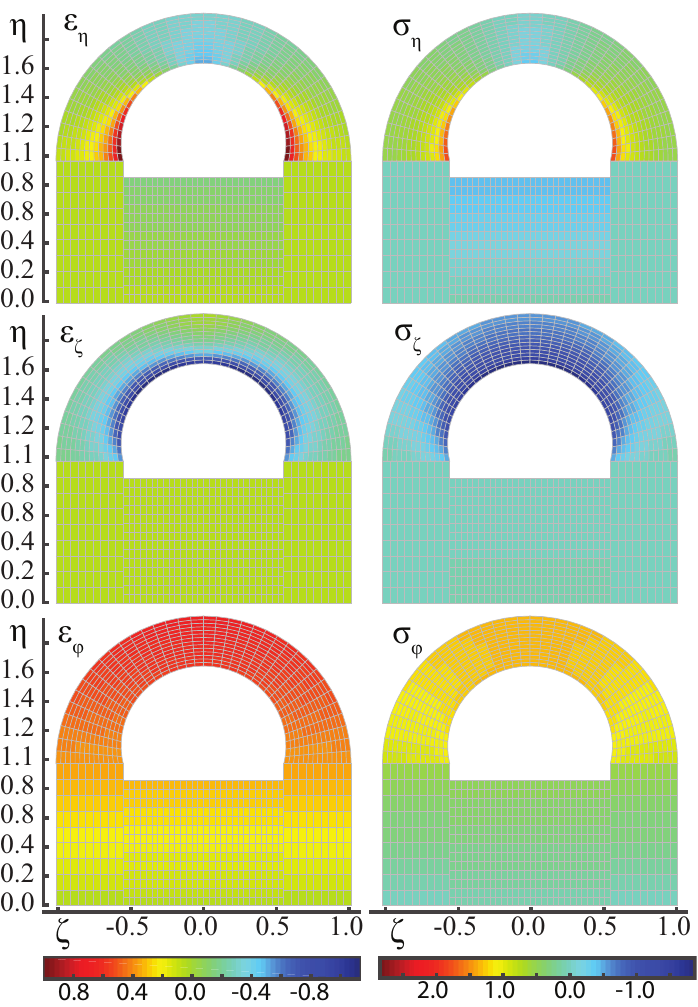}
\caption{Maps of stress and strain on a cross section. For the strains (left), we plotted $\varepsilon = \lambda - 1$, while for the stresses (right) we plotted the non-dimensional values $\sigma = \tilde{\sigma}/\mu$.}
\label{fig:S&S}
\end{figure}
We computed the maps of strains and stresses on the cross section of the structure. The results of the stresses and strains along the rectangular coordinates $\eta$ and $\zeta$ are shown in figure \ref{fig:S&S}.

As we can notice, due to the assumptions there are little discontinuities in some of the values of stresses and strains, mainly localized around the corners where the three sections touch each other.
As for the strains, it is interesting to notice that the maximum strains are all $\varepsilon \leq 1.2$, so $\lambda \leq 2.2$, thus confirming the formal equivalence in the use of the Neo-Hookean model respect to the Gent model, as it was assumed in section \ref{sec:mat_c}.
We can notice that the strain in longitudinal direction $\varepsilon_\varphi$, which is the one driving all the others, is homogeneous over all the cross-section and increases continuously with $\eta$.
As expected the points in the rectangular section are under compression along $\eta$, while they are under tension along $\phi$.

\begin{figure}[]
\centering
\includegraphics[width=\columnwidth]{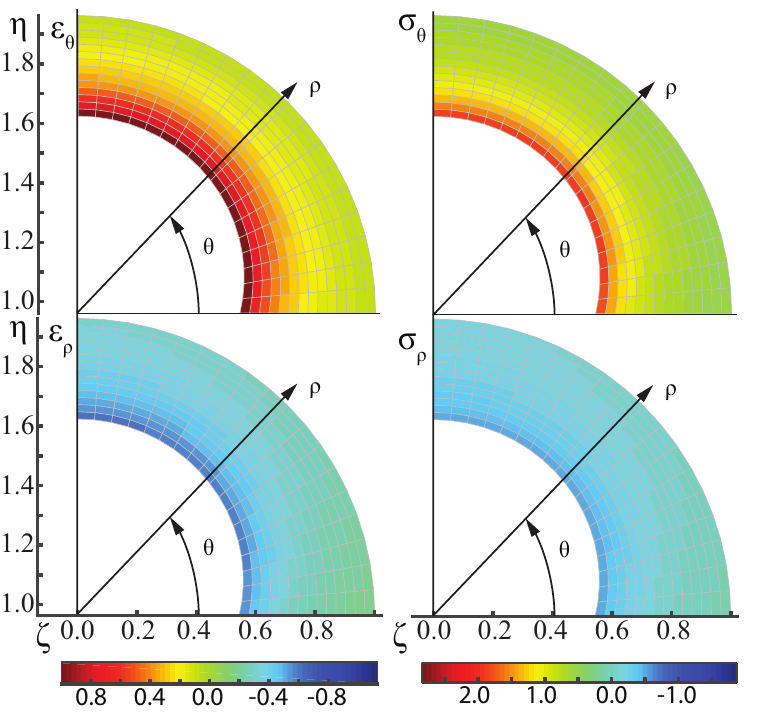}
\caption{Maps of polar stresses and strains on circular cross section. For the strains (left), we plotted $\varepsilon = \lambda - 1$, while for the stresses (right) we plotted the non-dimensional values $\sigma = \tilde{\sigma}/\mu$.}
\label{fig:S&S_circular}
\end{figure}
For a better understanding of the stresses and strains behaviours on the circular part of the cross section, we also plotted the $\rho$ and $\theta$ components in figure \ref{fig:S&S_circular}.
%
% We can observe that, coherently with the assumptions in section \ref{sec:kinc}, $\epsilon_\theta$ does not depend on $\theta$, while $\epsilon_\rho$ depends on both $\rho$ and $\theta$.
We can observe that, coherently with the assumptions of section \ref{sec:kinc}, $\varepsilon_\theta$ and $\varepsilon_\rho$ depend on both $\rho$ and $\theta$.
We have a state of compression along $\rho$ while a state of tension along $\theta$. In both cases the maximum values are localized at the internal radius $\rho=k(\theta)$. 

%Another important information is about the total level of stress in each point of the cross section. Using the distorsional energy density criterion (\cite{boresi1993advanced}), we computed the effective stress in terms of principal stresses as
%\begin{equation}
%\sigma_e = \sqrt[•]{0.5(\sigma_\eta - \sigma_\zeta)^2 + (\sigma_\zeta - \sigma_\phi)^2 + (\sigma_\phi - \sigma_\eta)^2 }
%\end{equation}

The results on the response of the actuator allow the functional design of the actuator. In facts, from the results in figure \ref{fig:validation} we can choose the value of static and dynamic pressures ($p_0$ and $\Delta p$) in order to obtain a certain target oscillation $\Phi_0 + \Delta \Phi$.
Additionally, by knowing the level of stress and strain we can design the structure in terms of materials and dimensions, fixed the value of the pressure and we can predict where the structure will eventually fail.

\subsection{Characterization of the actuator}
\label{sec:characterization}
%
%\begin{table}
%\centering
%\def\arraystretch{1.2}%  1 is the default, change whatever you need
%\caption{Parameters used in the $k^*$ characterization}
%\label{tab:ks} 
%\begin{tabular}{|c|c|c|c|c|}
%\hline
%\cellcolor[gray]{0.9}	& 
%$\mu$ [kPa] \cellcolor[gray]{0.9}	&
% $h^*$ [mm]  \cellcolor[gray]{0.9} &
%  $R_2^*$ [mm] \cellcolor[gray]{0.9} &
%    $ L^*$ [mm] \cellcolor[gray]{0.9}
%\\
%\hline
%1 \cellcolor[gray]{0.9} & 92 & 7.0 & 8.0 & 118
%\\
%\hline
%\end{tabular}
%\end{table}
%%

%
%
\begin{figure}[]
\centering
\includegraphics[width=0.9\columnwidth]{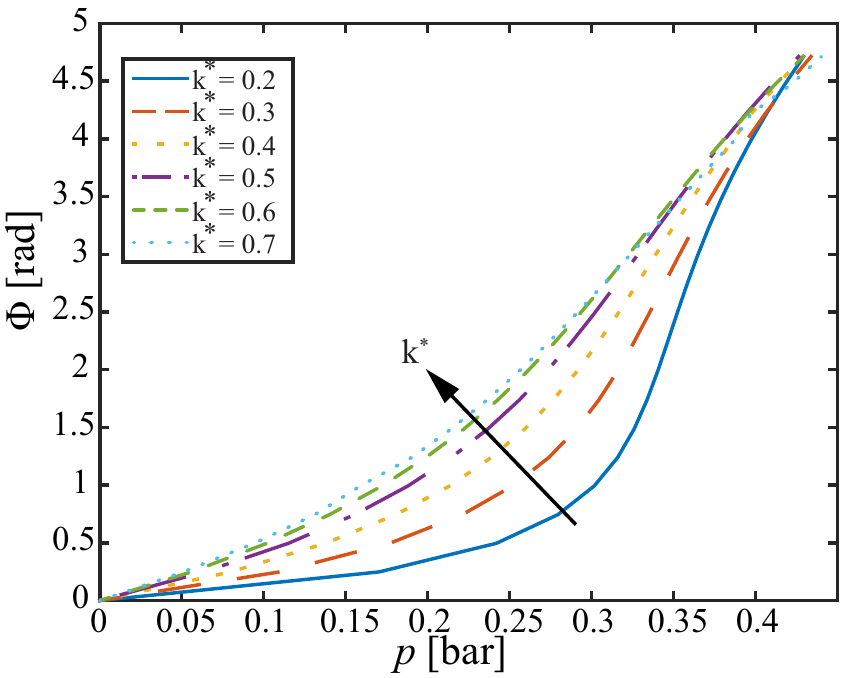}
\caption{Characterization of the actuator for different values of the ratio between internal and external radii in the undeformed configuration $k^*=R_1^*/R_2$.}
\label{fig:ks}
\end{figure}

We computed the angle-pressure characteristic for different values of the $k^*$ parameter in order to characterize the actuator. The parameters used for the computation correspond to the mock-up number 1 (table \ref{tab:mockup}) but with $R_1^*$ changing accordingly to $k^*$, while the results are shown in figure \ref{fig:ks}. As we can observe, the response of the actuator changes gradually by varying $k^*$, so it is possible to tune the response of the actuator by choosing the design parameter $k^*$.

%%
%%
%\begin{figure}[]
%\centering
%\includegraphics[width=\columnwidth]{Figures/q_role_k07.pdf}
%\caption{Characterization of the actuator for different values of $q^* = L^*/R_2$, representing the slenderness of the structure. Plot for $k^*=0.7$.}
%\label{fig:qs_k07}
%\end{figure}
%%
%%
%\begin{figure}[]
%\centering
%\includegraphics[width=\columnwidth]{Figures/q_role_k03.pdf}
%\caption{Characterization of the actuator for different values of $q^* = L^*/R_2$, representing the slenderness of the structure. Plot for $k^*=0.3$.}
%\label{fig:qs_k03}
%\end{figure}
%%
%%
%
\begin{figure}[]
\centering
\includegraphics[width=0.9\columnwidth]{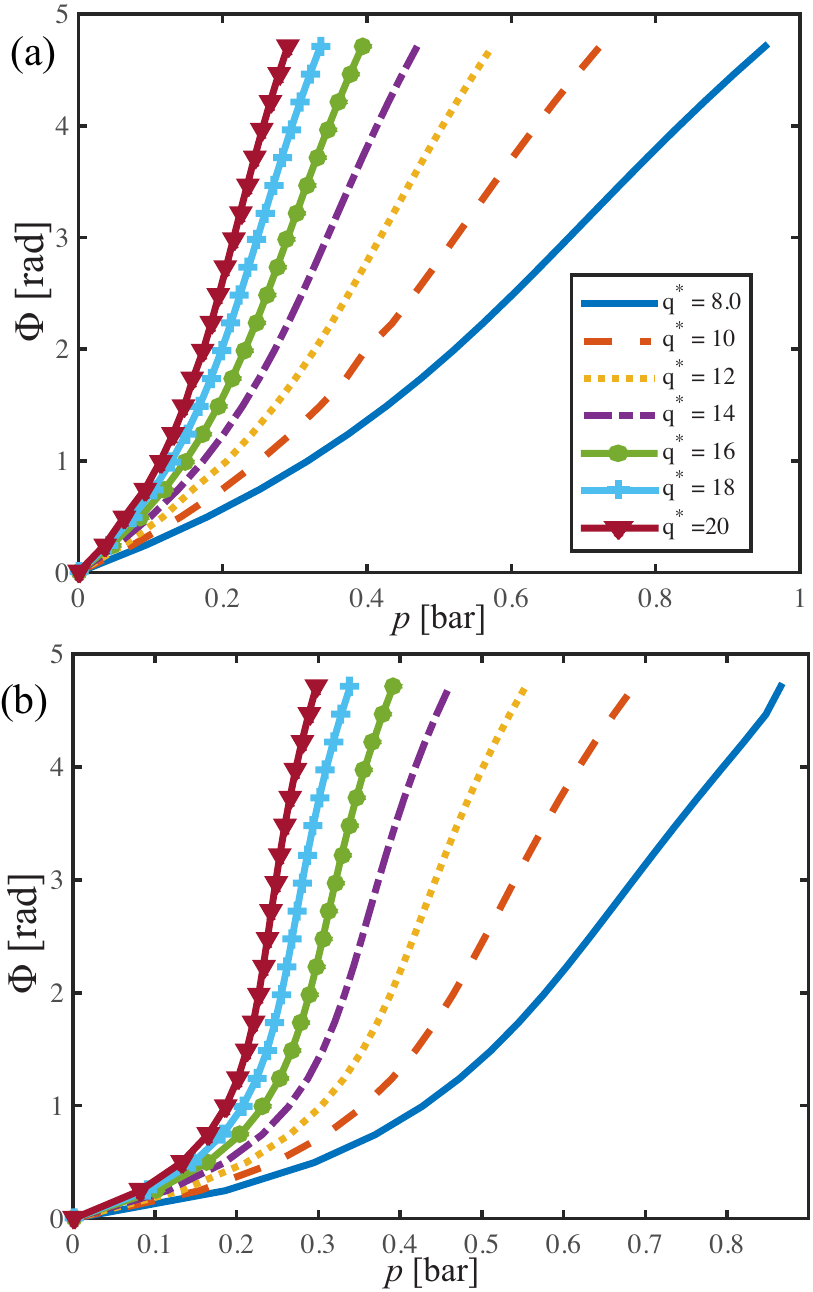}
\caption{Characterization of the actuator for different values of $q^* = L^*/R_2$, representing the slenderness of the structure. Plots for $k^*=0.7$ (a) and $k^*=0.3$ (b).}
\label{fig:qs_k07}
\end{figure}

We then chose two values of the ratio $k^*$ and analyzed the characterization for different values of $q^* = L^*/R_2$. The results are shown in figure \ref{fig:qs_k07}. The parameters are the same used before (table \ref{tab:mockup}, mock-up 1), but with $R_1^*$ and $L^*$ varying according to $k^*$ and $q^*$ respectively.
In this case, as expected, $q^*$ mainly influences the reactivity of the actuator, which results more steep for more slender structures ($q^*$ high) and more gradual for lower values of $q^*$.

Apart from the mechanical design of the actuator, the last two results obtained in this section are extremely useful for the functional design of the response of the actuator. In facts, being able to tune the nonlinear response of soft actuator is a highly beneficial feature, allowing the design of systems presenting snap-through instabilities to amplify the response \cite{Overvelde2015} or self-stabilized limit cycle oscillators \cite{slotine1991applied,cacucciolo2015adaptive}. On the other hand, by choosing a high value for $k^*$, which corresponds to a relatively thin wall thickness, the response can be close to linear.

\section{Conclusions}
\label{sec:conclusions}
In this work, we presented a quasi-static, 3D mechanical model for fibre-reinforced BFAs. The model includes the effect of the pressure on the lateral surface of internal chamber of the actuator and the non-constant torque induced by the pressure at the tip. All of the three principal strains and their interaction are taken into account.

In order to keep the complexity of the model at an acceptable level, we used the Eulerian description and applied all the equilibria in the a-priori defined deformed configuration.
The resulted formulation is readable and computationally efficient: it is possible to compute the kinematics once for all and store the results in a database; each step of the statics requires only numerical integrations of algebraic expressions.

We experimentally validated the model and then used it to reproduce the nonlinear response of these actuators and its variation with the design parameters.
This mechanical model can be useful in several ways: (1) it is a powerful tool for designing the actuator response for a target application; (2) can be used for the mechanical design of the structure (e.g., materials, dimensions); (3) by capturing the intrinsic mechanics of these systems, provides insights for the design of novel types of soft actuators.

\ack
This work is supported by: RoboSoft - A Coordination Action for Soft Robotics (FP7-ICT-2013-C \# 619319).
The authors would like to thank S. Maeda (Shibaura Institute of Technology, Tokyo, Japan) for valuable advice on material models.

%%%%% Begin the APPENDICES %%%%%%%

% \appendix

% \section{Diffusion constant}
% \label{app:DC}

%%%%%%%%%%%%%%%% Bibliography %%%%%%%%%%%
\section*{References}
\bibliographystyle{iopart-num}
\bibliography{/Users/VitoCacu/Dropbox/1_PhD/LIBRARY/library}

\providecommand{\newblock}{}
\begin{thebibliography}{10}
\expandafter\ifx\csname url\endcsname\relax
  \def\url#1{{\tt #1}}\fi
\expandafter\ifx\csname urlprefix\endcsname\relax\def\urlprefix{URL }\fi
\providecommand{\eprint}[2][]{\url{#2}}

\bibitem{maeda2015rapid}
Maeda S, Kato T, Kogure H and Hosoya N 2015 {\em Applied Physics Letters\/}
  {\bf 106} 171909 ISSN 0003-6951

\bibitem{shintake2016versatile}
Shintake J, Rosset S, Schubert B, Floreano D and Shea H 2016 {\em Advanced
  Materials\/} {\bf 28} 231----238 ISSN 09359648

\bibitem{Bufalo2008}
Bufalo G~D, Placidi L and Porfiri M 2008 {\em Smart Materials and Structures\/}
  {\bf 17} 045010 ISSN 0964-1726

\bibitem{Carpi2013}
Carpi F, Frediani G, Gerboni C, Gemignani J and {De Rossi} D 2013 {\em Medical
  Engineering {\&} Physics\/} {\bf 36} 205--211 ISSN 13504533

\bibitem{kim2013soft}
Kim S, Laschi C and Trimmer B 2013 {\em Trends in biotechnology\/} {\bf 31}
  287--294

\bibitem{laschi2014soft}
Laschi C and Cianchetti M 2014 {\em Frontiers in bioengineering and
  biotechnology\/} {\bf 2} 3 ISSN 2296-4185

\bibitem{majidi2014soft}
Majidi C 2014 {\em Soft Robotics\/} {\bf 1} 5--11

\bibitem{Polygerinos2014}
Polygerinos P, Wang Z, Galloway K~C, Wood R~J and Walsh C~J 2014 {\em Robotics
  and Autonomous Systems\/} {\bf 73} 135--143 ISSN 09218890

\bibitem{ranzani2015bioinspired}
Ranzani T, Gerboni G, Cianchetti M and Menciassi A 2015 {\em Bioinspiration
  {\&} biomimetics\/} {\bf 10} 035008 ISSN 1748-3190

\bibitem{Doumit2009}
Doumit M, Fahim a and Munro M 2009 {\em IEEE Transactions on Robotics\/} {\bf
  25} 1282--1291 ISSN 1552-3098

\bibitem{tolley2014resilient}
Tolley M~T, Shepherd R~F, Mosadegh B, Galloway K~C, Wehner M, Karpelson M, Wood
  R~J and Whitesides G~M 2014 {\em Soft Robotics\/} {\bf 1} 213--223

\bibitem{cacucciolo2015adaptive}
Cacucciolo V, Ansari Y, {Leylavi Shoushtari} A, Cianchetti M and Laschi C 2015
  {Adaptive locomotion on uneven terrains by means of a functional separation
  of time scales in the design and control of robots} {\em Proceedings of AMAM
  - Adaptive Motion in Animals and Machines\/}

\bibitem{deimel2014novel}
Deimel R and Brock O 2014 {\em Robotics: Science and Systems, Berkeley, CA\/}
  1687--1692

\bibitem{Martinez2013}
Martinez R~V, Branch J~L, Fish C~R, Jin L, Shepherd R~F, Nunes R~M~D, Suo Z and
  Whitesides G~M 2013 {\em Advanced Materials\/} {\bf 25} 205--212 ISSN
  09359648

\bibitem{Mosadegh2014}
Mosadegh B, Polygerinos P, Keplinger C, Wennstedt S, Shepherd R~F, Gupta U,
  Shim J, Bertoldi K, Walsh C~J and Whitesides G~M 2014 {\em Advanced
  Functional Materials\/} {\bf 24} 2163--2170 ISSN 16163028

\bibitem{renda2014dynamic}
Renda F, Giorelli M, Calisti M, Cianchetti M and Laschi C 2014 {\em Robotics,
  IEEE Transactions on\/} {\bf 30} 1109--1122 ISSN 1552-3098

\bibitem{bishop-moser2015design}
Bishop-moser J and Kota S 2015 {\em Robotics, IEEE Transactions on\/} {\bf 31}
  536--545

\bibitem{Sorge2015}
Sorge F 2015 {\em Meccanica\/} {\bf 50} 1371--1386 ISSN 0025-6455

\bibitem{polygerinos2015modeling}
Polygerinos P, Wang Z, Overvelde J~T~B, Galloway K~C, Wood R~J, Bertoldi K and
  Walsh C~J 2015 {\em Robotics, IEEE Transactions on\/} {\bf 31} 778--789

\bibitem{ogden1997nonlinear}
Ogden R~W 1997 {\em {Non-linear elastic deformations}\/} (Courier Corporation)

\bibitem{Overvelde2015}
Overvelde J~T~B, Kloek T, D'haen J~J~A and Bertoldi K 2015 {\em Proceedings of
  the National Academy of Sciences of the United States of America\/} {\bf 112}
  10863--10868 ISSN 1091-6490

\bibitem{sprowitz2013towards}
Spr{\"{o}}witz A, Tuleu A, Vespignani M, Ajallooeian M, Badri E and Ijspeert
  A~J 2013 {\em The International Journal of Robotics Research\/} {\bf 32}
  932--950 ISSN 0278-3649, 1741-3176

\bibitem{buchanan2005analysis}
Buchanan G~R and Liu Y~J 2005 {\em International Journal of Mechanical
  Sciences\/} {\bf 47} 277--292 ISSN 00207403

\bibitem{Gent1996}
Gent A~N 1996 {\em Rubber Chemistry and Technology\/} {\bf 69} 59--61 ISSN
  0035-9475 (\textit{Preprint} \eprint{0002163820})

\bibitem{More1983}
Mor{\'{e}} J~J and Sorensen D~C 1983 {\em SIAM Journal on Scientific and
  Statistical Computing\/} {\bf 4} 553--572 ISSN 0196-5204

\bibitem{Steihaug1983}
Steihaug T 1983 {\em SIAM Journal on Numerical Analysis\/} {\bf 20} 626--637
  ISSN 0036-1429

\bibitem{slotine1991applied}
Slotine J~J~E and Li W 1991 {\em {Applied nonlinear control}\/} (Prentice-hall
  Englewood Cliffs, NJ)

\end{thebibliography}

\end{document}